\documentclass[runningheads]{llncs}

\usepackage{graphicx}
\usepackage{subfigure}

\usepackage{amsfonts} 
\usepackage{url}
\usepackage{color}
\usepackage[sort,nocompress]{cite} 
\usepackage[table,xcdraw]{xcolor}
\usepackage{amsfonts} 
\usepackage{amsmath} 
\usepackage{tikz}
\usetikzlibrary{arrows}

\usepackage{hyperref}

\usepackage{listings}

\usepackage[ruled,linesnumbered,noend]{algorithm2e}

\begin{document}
\title{Graph-based process mining}

\author{Amin Jalali}

\institute{
	Department of Computer and Systems Sciences\\
	Stockholm University, Sweden \\
	\email{aj@dsv.su.se}
}

\authorrunning{A. Jalali}
\titlerunning{Graph-based process mining}

\maketitle

\begin{abstract}
Process mining is an area of research that supports discovering information about business processes from their execution event logs. The increasing amount of event logs in organizations challenges current process mining techniques, which tend to load data into the memory of a computer. This issue limits the organizations to apply process mining on a large scale and introduces risks due to the lack of data management capabilities.
Therefore, this paper introduces and formalizes a new approach to store and retrieve event logs into/from graph databases. It defines an algorithm to compute Directly Follows Graph (DFG) inside the graph database, which shifts the heavy computation parts of process mining into the graph database. Calculating DFG in graph databases enables leveraging the graph databases' horizontal and vertical scaling capabilities in favor of applying process mining on a large scale. Besides, it removes the requirement to move data into analysts' computer. Thus, it enables using data management capabilities in graph databases.
We implemented this approach in Neo4j and evaluated its performance compared with current techniques using a real log file. The result shows that our approach enables the calculation of DFG when the data is much bigger than the computational memory. It also shows better performance when dicing data into small chunks.

\vspace{+3pt}
\textbf{Keywords:} Process mining, graph database, Big Data, Neo4j

\end{abstract}

\section{Introduction}\label{Sec:Introduction}

Business Process Management (BPM) is a research area that aims to enable organizations to narrow the gap between business goals and information technology support~\cite{WeskeBPM2019}. 
Business process evaluation is a key support in narrowing down this gap. 
There are two evaluation techniques to analyze business processes, a.k.a., model-based analysis, and data-based analysis~\cite{AalstBPMSurvey2013}. 
While model-based analysis deals with the analysis of business process models, the data-based analysis mostly focuses on analyzing business processes based on their execution event logs. 

Process Mining is a discipline in the BPM area that enables data-based analysis for business processes in organizations~\cite{AalstPMBook2016}. 
It allows analysts not only to evaluate the business processes but also to perform process discovery, compliance checking, and process enhancement based on the execution result, a.k.a., event logs. 
As the volume of logs increases, new opportunities and challenges also appear. 
The large volume of logs enables the discovery of more information about business processes; while also raises some challenges, such as feasibility, performance, and data management.

Most process mining techniques require data to be loaded first into memory, which is a feasibility technical challenge when applying them on a large volume of data in a single computer. 
They also need to work on fine-grain data that might not be accessible to analysts due to organizations' data management policies, which is an organizational challenge. 
Most of the organizations apply many restrictions to grant analysts access in the granular level to data, which can hinder applying process mining techniques. 
These are some challenges that the author also faced when applying process mining in practice.

To address these challenges, this paper proposes and formalizes a new approach to store and retrieve event logs in graph databases. 
It also defines an algorithm to compute Directly Follows Graph (DFG) inside the graph database, which shifts the heavy computation parts of process mining into the graph database. 
As a result, it enables i) removing the requirement to move data into analysts' computer, ii) applying graph databases' fine-grained access control on event logs and preserving privacy while applying process mining, and iii) scaling the DFG computation vertically and horizontally.

The approach is implemented in Neo4j, and its performance is evaluated in comparison with current techniques based on a real log file. 
The result shows that the approach can discover process models when the data is much bigger than the computational memory. 
It also shows better performance when dicing data into small chunks. 

The remainder of this paper is organized as follows.
Section~\ref{Sec:Background} gives a short background on process mining and graph database.
Section~\ref{Sec:Approach} introduces the graph-based process mining approach, and 
Section~\ref{Sec:Implementation} elaborates on the implementation of the approach in Neo4j.
Section~\ref{Sec:Evaluation} reports the evaluation results.
Section~\ref{Sec:RelatedWork} discuss alternative approaches and related works, and finally, 
Section~\ref{Sec:Conclusion} concludes the paper and introduces future research.

\section{Background}\label{Sec:Background}


\subsection{Process Mining}

Process Mining is a research area that supports business process data-based analysis.
The process mining approaches can be categorized as discovery, conformance, and enhancement~\cite{AalstPMBook2016}. 
The process discovery topic enables producing process models from event logs. The conformance topic enables checking the conformance of event logs with existing process models. Process enhancement topic enables improving process models using identified aspects from event logs. 
In all these topics, event logs are essential to enable process mining.

\begin{figure}[t!] 
	\begin{center}
		\includegraphics[width=1\textwidth]{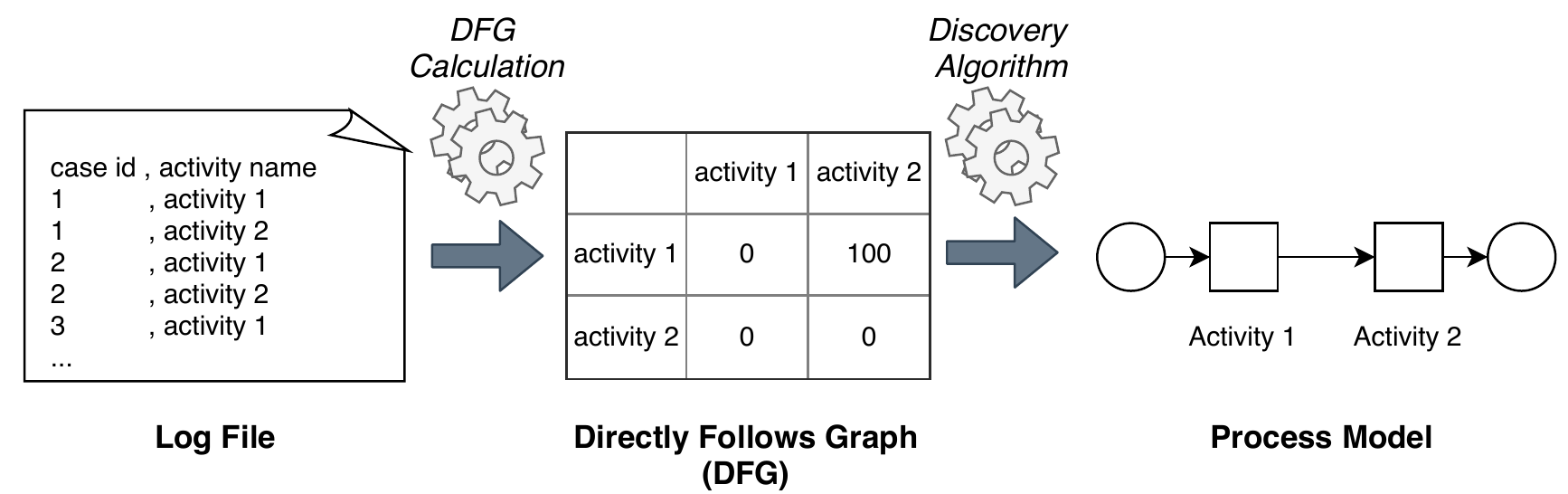}
		\vspace{0\baselineskip}
		\caption{Steps in a process discovery algorithm}
		\label{Fig:ProcessMining}
		\vspace{-2\baselineskip}
	\end{center}
\end{figure}

\figurename~\ref{Fig:ProcessMining} shows an overview of common process discovery steps, which can also be followed in conformance and enhancement as well.
The process discovery usually includes loading the log file, calculating the Directly Follows Graph (DFG), and discovering the process model from DFG.

The process discovery starts by loading a log file that stores business process execution results, a.k.a., log files. 
Each log contains a set of traces representing different cases that performed in the business process.
Each trace contains a set of events representing the execution result of activities in the business process. 
Thus, a log file shall contain information about traces and events at a minimum.
Note that with this basic set up, the events should be stored according to the execution order, unless we have information about execution time. 
It is usual to have more information like the execution time and resource who has done the activity on the log as well.

The next step is calculating the Directly Follows Graph (DFG). 
This graph shows the frequency of direct relations between activities that are captured in the log file.
The result can be considered as a matrix with the activity names at rows and columns. 
Let's consider the cell that has the row for \textit{activity 1} and columns for \textit{activity 2} in \figurename~\ref{Fig:ProcessMining}. 
The number in the cell shows the number of times that the \textit{activity 2} happened after \textit{activity 1}. 
Although the calculation of DFG comes back to alpha miner, which introduced around 20 years ago, it is still the backbone for many process mining algorithms and tools~\cite{AalstReview2020}. 
There are different variations of DFG that store more information, but the basic idea is the same.

Calculating DFG can be very time consuming and costly if the log file is huge. 
Also, many algorithms which are implemented in different tools like ProM and PM4PY requires to load logs to the memory first, which is problematic when calculating the big log files. 
The process mining algorithms also need to work with the most granular level of data, i.e., events, to calculate DFG. 
Thus analysts can face data management and security issues when discovering models, where companies might not be interested in handing over the complete set of data. 
These limitations can hinder applying process mining, while analysts are only interested in discovering the big picture rather than investigating individual cases. 
We will explain how our approach will solve this problem by enabling the calculation of DFG without granting access to event data in the Section~\ref{Sec:Approach}.

The last step is to infer the process model from DFG based on rules that are specified by a process discovery algorithm. 
This step usually does not take much time since the computation is performed on top of DFG.

\subsection{Graph Database}

Graph databases are Database Management Systems (DBMS) that support creating, storing, retrieving, and managing graph database models. 
Graph database models are defined as the data structure where the schema and instances are modeled as graphs, and the operation on graphs are graph-oriented~\cite{AnglesGraphSurvey2008}.
The idea is not new, and it comes back to the late eighties when the object-oriented models were also introduced~\cite{AnglesGraphSurvey2008}. 
However, it recently got much attention in both research and industry due to its ability to handle the huge amount of data and networks. 
It enables leveraging parallel computing capabilities to analyze massive graphs. 
As a result, a new discipline is emerged in research, called Parallel Graph Analytics~\cite{lenharthParallelGraph2016}.

There are different sorts of graph databases with different features. 
For example, Neo4j is a graph DBMS that supports both vertical and horizontal scaling, meaning that not only the hardware of the system that runs the DBMS can be scaled out but the number of physical nodes that run the DBMS as a network can be increased. 
These features enable having a considerable performance at runtime. 
Also, it allows different sorts of access controls, such as Fine-grained access control, Sub-graph access control, and role-based access control. 
In this way, different data management strategy can be applied. 

\begin{figure}[b!] 
	\begin{center}
		\includegraphics[width=1\textwidth]{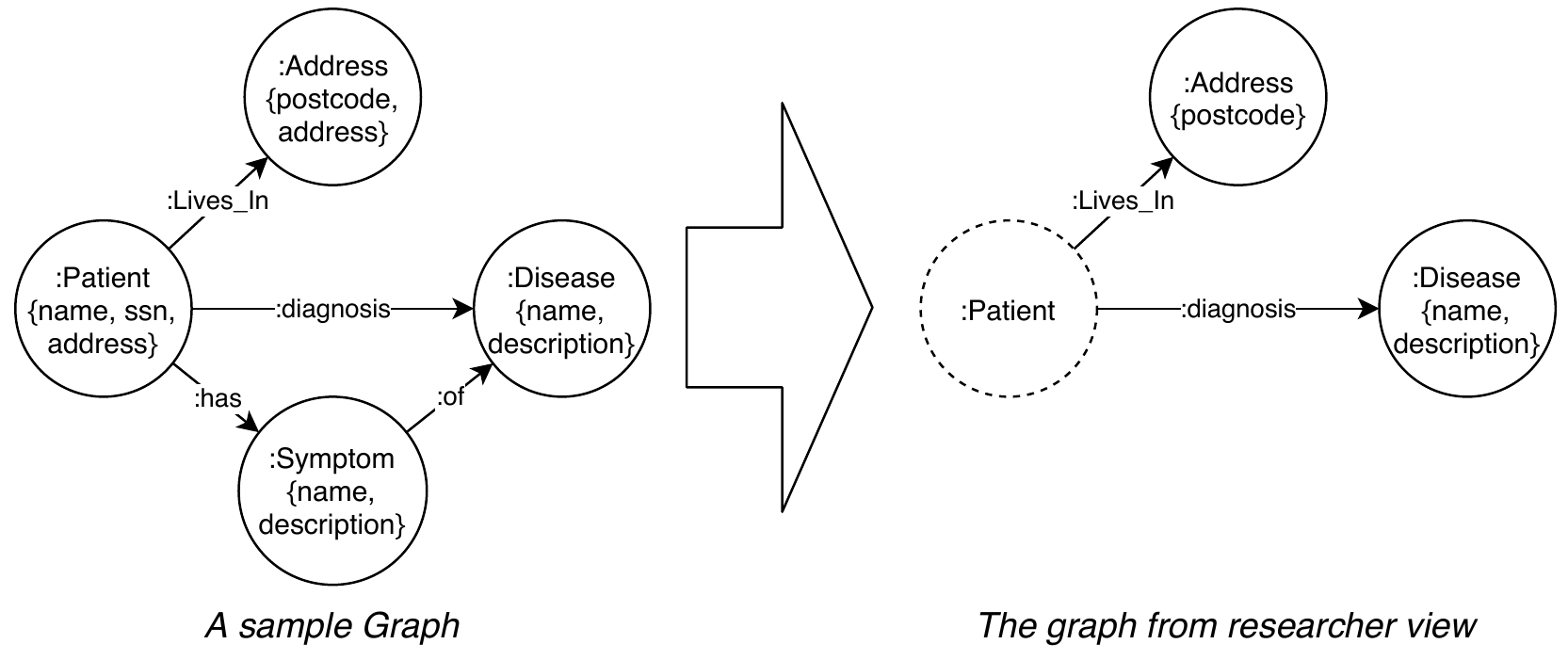}
		\vspace{0\baselineskip}
		\caption{An example of a graph schema in a graph database}
		\label{Fig:GraphIntro}
		\vspace{-2\baselineskip}
	\end{center}
\end{figure}

The left-side of \figurename{\ref{Fig:GraphIntro}} shows a fictitious short example schema for a simple graph in a graph database. 
It has four nodes, i.e., patient, disease, symptoms, and address. 
It also has relations among the nodes that specify their connection. 
For example, a patient can live in an address, and a patient can be diagnosed with a disease. 
Each node can have properties, e.g., the patient has a social security number (ssn). 

In a graph database, it is possible to grant access to analysts to analyze the spread of diseases, e.g., COVID19, based on patients who live in the same address but not reveal the patient's information nor individual address. 
We can, as an example, grant access to the postal code level.
To do so, we can set the access in a way that analysts can traverse relations from/to patients but cannot see patient's information. 
Also, we can hide individual addresses and symptoms.
Thus, the schema will be like the right-side of \figurename{\ref{Fig:GraphIntro}} for analysts. 
In this way, the researcher can analyze how people who lived in the same address got the same disease.
Also, the result can be shown in the postal code level.
The build-in access control mechanism enables us to design an approach that preserves the privacy of resources while performing process mining.

To traverse or query the graph, graph databases define their query language. Cypher~\cite{Cypher2018} is a declarative query language that allows the application of graph operations on graph databases, which is implemented in several graph databases, including Neo4j.

\section{Approach}\label{Sec:Approach}


This section formalizes the approach and explains it through an example. 
We simplify the formal definition by limiting the set of attributes to hold information about activities. 
In practice, the definition of attributes can be extended to store all information about the data perspective.
The approach enables preserving the privacy of resources while computing DFG. 
This requires defining logs, traces, events, and event attributes as different nodes so that the access control can be limited for each node.

\subsection{Definitions}

\begin{definition}[Event Repository]
	An event repository is a tuple $G=(N=L\cup T\cup E\cup A, R)$, where:
	\begin{itemize}
		\item $N$ is the set of nodes with the following subsets:
			\subitem $L$ represents the set of logs,
			\subitem $T$ represents the set of traces,
			\subitem $E$ represents the set of events,
			\subitem $A$ represents the set of attributes, representing activities, where:
			\subitem $L\cap T\cap E\cap A=\emptyset$.
		\item $R=L\times T\cup T\times E\cup E\times E\cup E\times A$ is the set of relations connecting:
			\subitem logs to traces, i.e., $L\times T$ 
			\subitem traces to events, i.e., $T\times E$,
			\subitem events to events, i.e., $E\times E$,
			\subitem events to attributes, i.e., $E\times A$, where:
			\subitem $N\cap R=\emptyset$
	\end{itemize}
	Let's also define two operators on the graph's nodes as:
	\begin{itemize}
		\item $\bullet n$ represents the operator that retrieves the set of nodes from which there are relations to node $n$, i.e., $\bullet n=\{\forall e\in N| (e,n)\in R\}$.
		\item $n\bullet$ represents the operator that retrieves the set of nodes to which there are relations from node $n$, i.e., $n\bullet=\{\forall e\in N| (n,e)\in R\}$.
	\end{itemize}
\end{definition}

\begin{definition}[Soundness]
	An event repository $G=(N=L\cup T\cup E\cup A, R)$, where $N, L, T, E, A, R$, representing the set of Nodes, Logs, Traces, Events, Attributes, Relations respectively, is sound iff:
	\begin{itemize}
		\item $\forall t\in T, |\bullet t|=1 $, meaning that a trace must belongs to 1 and only 1 log.
		\item $\forall e\in E, |\bullet e \cap T|=1 $, meaning that an event must belongs to 1 and only 1 trace.
		\item $\forall e\in E, |\bullet e \cap E|<=1 $, meaning that an event can only have at most 1 input flow from another event.
		\item $\forall e\in E, |e\bullet \cap E|<=1 $, meaning that an event can only have at most 1 output flow to another event.
		\item $\forall e\in E, |e\bullet \cap A|=1 $, meaning that an event must be related to 1 and only 1 attribute.
	\end{itemize}
\end{definition}


Note that this formalization can be extended to enable several types of sequences among event logs.
To calculate DFG, we need to count the number of direct relations among events for each activity pairs. 
Algorithm 1 defines how the DFG for a given sound event repository can be calculated.

\begin{algorithm}[h!]\label{Algorithm1}
	\SetAlgoLined 
	\SetKwFunction{algo}{dfgcalculator}
	
	\SetKwProg{myalg}{Algorithm}{}{}
	\myalg{\algo{$G=(N=L\cup T\cup E\cup A, R)$}}{
		$\Psi \leftarrow \emptyset$\;
		\ForEach {two attributes $a,b \in A$}{
			$c=\mathlarger{\sum_{\forall e\in\bullet a, e^\prime\in\bullet b}{|(e,e^\prime)\in R|}}$\;
			$\Psi \leftarrow \Psi \cup \{(a,b,c)\}$;
			}
		}
		\KwRet $\Psi$\;{}
	\caption{Algorithm for calculating dfg}
\end{algorithm} 


\subsection{Example}

This section elaborates on the definitions through an example.

\figurename{\ref{Fig:SampleGraphRepository}} shows an example of a sound event repository graph. 
The set of nodes for Log, Trace, Event, and Attribute are collored as green, red, white, and yellow respectively. 
This repository includes one log file, called \textit{l1}, which has two traces, i.e., \textit{t1} and \textit{t2}. 
\textit{t1} has three events that are happened with this order \textit{e1} $\rightarrow$ \textit{e2} $\rightarrow$ \textit{e3}. 
\textit{t2} also has three events that are happened with this order \textit{e4} $\rightarrow$ \textit{e5} $\rightarrow$ \textit{e6}. 

As it can be seen, each event is related to one activity, e.g., \textit{e1} is the execution of activity \textit{a1}. 
To get the list of events that happened for an activity \textit{a1}, we can use $\bullet a1$ operator, which returns $\{e1\}$. 
For some activities, there might be more than one events, e.g., $\bullet a2$ returns $\{e2, e4\}$.
Applying Algorithm 1 on this event repository will return the DFG. 
The DFG calculation is described as bellow:

\begin{itemize}
	\item for each pair of activities, the algorithm will calculate the frequency. We show the calculation for one pair example, i.e., $a2,a3$:
	\subitem $\bullet a2$ retreives $\{e2, e4\}$
	\subitem $\bullet a2$ retreives $\{e3, e5\}$
	\subitem $c=\mathlarger{\sum_{\forall e\in\bullet a2, e^\prime\in\bullet a3}{|(e,e^\prime)\in R|}}=\mathlarger{\sum_{\forall e\in\{e2, e4\}, e^\prime\in\{e3, e5\}}{|(e,e^\prime)\in R|}}\\=|\{(e2,e3),(e4,e5)\}|=2$\;
\end{itemize}

\definecolor{log}{rgb}{.5,0.9,0.5}
\definecolor{trace}{rgb}{.9,0.5,0.5}
\definecolor{event}{rgb}{1,1,1}
\definecolor{attribute}{rgb}{.9,0.9,0}

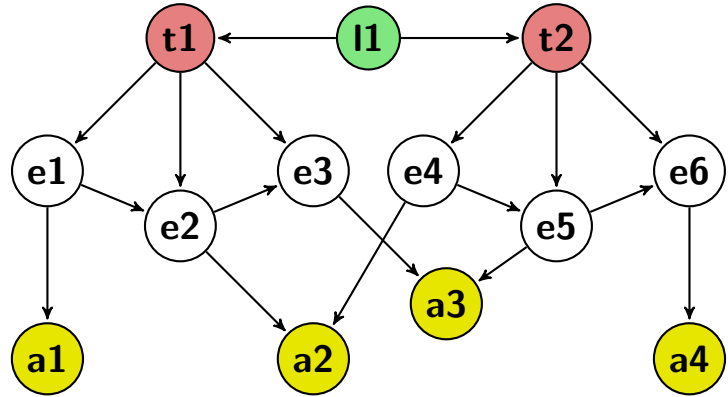
\begin{figure}[t!]
	\centering
	\begin{tikzpicture}[->,>=stealth',shorten >=1pt,auto,node distance=2.5cm,
	thick,main node/.style={circle,draw,font=\sffamily\Large\bfseries}]
	\node[main node,fill=log] (1) {l1};
	\node[main node,fill=trace] (2) [left of=1] {t1};
	\node[main node,fill=event] (3) [below left of=2] {e1};
	\node[main node,fill=event] (4) [below of=2] {e2};
	\node[main node,fill=event] (5) [below right of=2] {e3};
	\node[main node,fill=trace] (6) [right of=1] {t2};
	\node[main node,fill=event] (7) [below left of=6] {e4};
	\node[main node,fill=event] (8) [below of=6] {e5};
	\node[main node,fill=event] (9) [below right of=6] {e6};
	\node[main node,fill=attribute] (10) [below of=3] {a1};
	\node[main node,fill=attribute] (11) [below right of=4] {a2};
	\node[main node,fill=attribute] (12) [below right of=5] {a3};
	\node[main node,fill=attribute] (13) [below of=9] {a4};
				
	\path[every node/.style={font=\sffamily\small}]
	(1) edge node {} (2)
	(2) edge node {} (3)
	(2) edge node {} (4)
	(2) edge node {} (5)	
	(3) edge node {} (4)
	(4) edge node {} (5)
	(1) edge node {} (6)
	(6) edge node {} (7)
	(6) edge node {} (8)
	(6) edge node {} (9)	
	(7) edge node {} (8)
	(8) edge node {} (9)
	(3) edge node {} (10)
	(4) edge node {} (11)
	(5) edge node {} (12)
	(7) edge node {} (11)
	(8) edge node {} (12)
	(9) edge node {} (13)
	;
	\end{tikzpicture}
	\caption{An example of a sound event repository graph} \label{Fig:SampleGraphRepository}
\end{figure}

If we calculate all possibilities, the result will be like \tablename{\ref{Table:DFG}}.

\begin{table}[h!]
	\centering
	\begin{tabular}{|l|l|l|l|l|}
		\hline
		& a1 & a2 & a3 & a4 \\ \hline
		a1 & 0  & 1  & 0  & 0  \\ \hline
		a2 & 0  & 0  & 2  & 0  \\ \hline
		a3 & 0  & 0  & 0  & 1  \\ \hline
		a4 & 0  & 0  & 0  & 0  \\ \hline
	\end{tabular}
	\vspace{1\baselineskip}
	\caption{DFG calculation for the sample event repository graph} \label{Table:DFG}
\end{table}


\section{Implementation}\label{Sec:Implementation}

The approach presented in this paper is implemented in Neo4j, which was chosen because it supports 
i) storing graphs and doing graph operations, 
ii) both vertical and horizontal scaling, 
iii) querying the graph using Cypher,
iii) fine-grained access control, sub-graph access control, and role-based access control. 
The supported access controls enable analysts to analyze event logs without the need to have access to individual details.


We implemented a data-aware version of the approach. 
The main differences with the formalization are:
\begin{itemize}
	\item Attributes have key and val, where key is set to predefined values, which are common in process mining applications, i.e., 'log\_concept\_name', 'case\_concept\_name', and 'concept\_name' representing the attribute that holds information for log, case, and activity name respectively.
	\subitem This feature means that the activities are stored as one subset of activity set, and they shall be filtered based on the attributes' data.
	\item events have timestamps to enable dicing information based on time. Note that the timestamp cannot be defined as an attribute with its own key since we will end up with many extra nodes due to many timestamps that exist for each event. Thus, they are kept as an attribute of Event class, following the same practice to deal with times in data warehousing~\cite{kimball2011data}.
\end{itemize}

The calculation of DFG is implemented using a Cypher query as bellow:

\lstset{frame=tb,
	basicstyle={\small\ttfamily},
	breaklines=true,
}

\begin{lstlisting}
match 
(a1:Attribute {key:'concept_name'})<--(:Event)-[n]->(:Event) -->(a2:Attribute {key:'concept_name'})
return 
a1.val as dfg_from, a2.val as dfg_to, count(n) as dfg_freq
\end{lstlisting}

The match clause in the query identifies all patterns in sub-graphs that match the expression. 
The expression select two attributes \textit{a1} and \textit{a2} with the type of \textit{concept\_name}, which indicates that they are activities' names. 
Then, it selects all incoming events to those attributes where there is a direct relationship between those two events. 
The return clause retrieves all combinations of attributes in addition to the number of total direct relations between their events, which is the calculation that we formalized in Algorithm 1. 

To limit the number of events base on their timestamp, we can easily add a where clause to the cypher query to limit the timestamp. 
For other attributes, the associated attribute node can be filtered.
\section{Evaluation}\label{Sec:Evaluation}

This section reports the evaluation result, which is done by calculating DFG using our approach and process mining for python (pm4py) library~\cite{berti2019process} based on a real public log file~\cite{dees2016bpi}. 
This dataset is selected because it is published openly, which makes the experiment repeatable.
It is also the biggest log file that we could find in the BPI challenges, which can help us to evaluate the performance.

To evaluate the performance, we need to control the resources that are available for performing process mining. 
Thus, we decided to containerize the experiments and run them with Docker. 
Docker is a Platform as a Service (PaaS) product that enables creating, running, and managing containers. 
It also enables the control of the resources that are available for each container, such as RAM and CPU.

Among different process mining tools, we chose process mining for python (pm4py) library~\cite{berti2019process}, because i) it is open-source; ii) the DFG calculation step and discovery step can be separated easily, and iii) it can easily be encapsulated in a container.
The separation of DFG calculation and discovery step in this library also enables reusing all discovery algorithms along using our approach, which makes our approach very reusable.

We designed two experiment to evaluate our approach, which are listed in \tablename{\ref{TBL:EvaluationSetting}}\footnote{The data and code can be found at \url{https://github.com/jalaliamin/neo4jpm}.}. 
In \textit{Experiment 1}, we loaded the whole log file into both containers running neo4j and pm4py, so we kept the number of event logs constant. 
We calculated DFG several times by changing the RAM and CPU, so we defined the computational resources as a variable.
In \textit{Experiment 2}, we kept RAM and CPU constant for both containers, and we calculated DFG by dicing the data. The dicing is done based on a time constraint, and we added more days in an accumulative way to increase the number of events.
We ran the experiments for each container separately to make sure that the assigned resources are free and available.

\begin{table}[t!]
	\begin{center}
		\begin{tabular}{|l|l|l|}
			\hline
			& Constant   & Variable   \\ \hline
			Experiment 1 & Events in the Log & CPU \& RAM \\ \hline
			Experiment 2 & CPU \& RAM & Events in the Log \\ \hline
		\end{tabular}
		\label{TBL:EvaluationSetting}
		\vspace{1\baselineskip}
		\caption{Evaluation setting}
	\end{center}
\end{table}

\subsection{Experiment 1}

To simulate the situation where the computational memory is less than the log size, we started by assigning 512 megabytes of ram to each container.
We added the same amount of RAM in each experiment round until we reached 4 gigabytes. 
We also changed the CPU starting from half of a CPU (0.5), adding by the same amount in each round until we reached 4.0.

\begin{figure}[b!] 
	\begin{center}
		\includegraphics[width=1\textwidth]{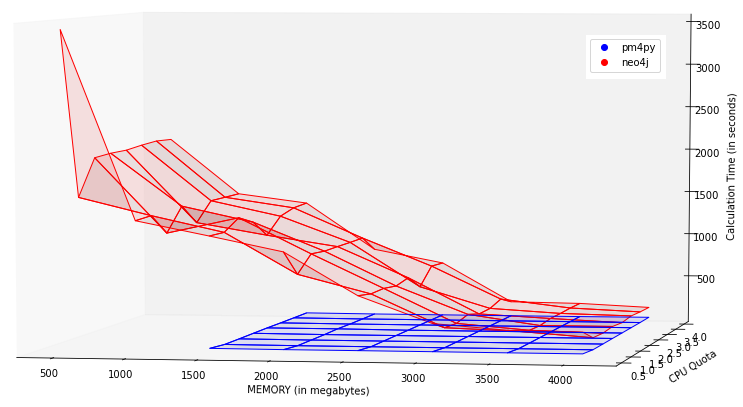}
		\vspace{0\baselineskip}
		\caption{Evaluating DFG calculation time by scaling resources}
		\label{Fig:Experiment1}
		\vspace{-2\baselineskip}
	\end{center}
\end{figure}

\figurename{\ref{Fig:Experiment1}} shows the execution result for both containers, where the
x, y and z axes refer to the available memory (RAM) (in megabytes), DFG calculation time (in seconds), and available CPU quotes, respectively.
The experiment related to neo4j and pm4py containers is plotted by red and blue, respectively.
As can be seen, pm4py could not compute DFG when the memory was less than the size of the log, i.e., around 1.5 gigabytes, while neo4j could calculate DFG in that setting. 
This shows that the graph database can compute DFG when computational memory is less than the log size, which is an enabler when applying process mining on a very large volume of data.

As it can be seen in the figure, the increasing amount of memory reduced the time that neo4j computed the DFG, while it has very little effect on pm4py. 
This is no surprise for in-memory calculation since if the log fits the memory, then the performance will not be increased much by adding more memory. 
It is also visible that assigning more CPU does not affect the performance of either of these approaches.

It should also be mentioned that despite increasing memory can reduce the DFG calculation time for neo4j significantly; it cannot be faster than pm4py when calculating the DFG on the complete log file. 
The reason can be that graph databases shall process metadata, which adds more computation than in-memory calculation approaches. 
Thus, for small log files that can fit the computer's memory, the in-memory approach can be better if the security and access control are not necessary.

\subsection{Experiment 2}

Event logs usually contain different variations that exist in the enactment of business processes~\cite{WilCubeCase2015}. These variations make process mining challenging because discovering the process based on the whole even logs usually produces so-called spaghetti models, which usually cannot be comprehended by humans, so they have very little value. Thus, analysts need to filter data to produce a meaningful model, which is a common practice in applying process mining~\cite{WilCubeCase2015,JalaliBPIC2016}. Therefore, we designed this experience to compare our approach and pm4py when calculating DFG on a filtered subset of data.

\begin{figure}[t!] 
	\begin{center}
		\includegraphics[width=0.7\textwidth]{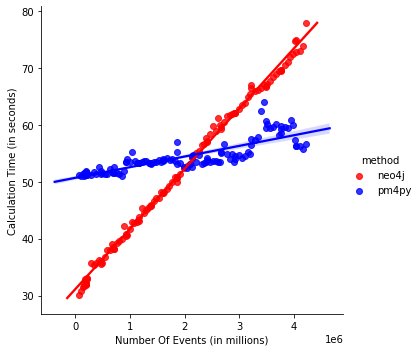}
		\vspace{0\baselineskip}
		\caption{Evaluating DFG calculation time by dicing the log}
		\label{Fig:Experiment2}
		\vspace{-2\baselineskip}
	\end{center}
\end{figure}

To evaluate this scenario, we kept the resources (RAM and CPU) constant for both containers, but we changed the condition based on which the data was filtered. 
This means that we kept the number of events in the log as a variable.
We assigned 14 Gb for RAM and 4 CPU for each container, which was run separately. 
We diced the data in both settings by filtering events that happened during the first day; then, we added one more day to the filter condition to increase the events in an accumulative way. 
We repeated this step for almost four months. 
In this way, we could compare the performance by considering how the size of the filtered events affects the performance of calculating DFG.

\figurename{\ref{Fig:Experiment2}} shows the evaluation result, where the x and y axes refer to the number of events (in millions) and DFG calculation time (in seconds). 
As can be seen, our approach performed better when the number of events is less than 2 million. 
Note that this is still a very big sub-log to analyze for process mining, so this shows that our approach can improve the performance of process mining when dealing with sub-set of the log. 
However, pm4py performed better when the number of events exceeded 2 million. 
This is no surprise since pm4py loaded logs into memory first, so increasing the size will have less effect on its performance. 
Indeed, the difference is only related to filtering the log and retrieving the biggest chunk of data in each iteration.

\section{Related Work and Discussion}\label{Sec:RelatedWork}

This section discusses the relationship between our approach to other research work.
The related work can be divided into three categories reflecting those that are about scalability, preserving privacy, and using graph databases in process mining.

\subsection{Scalability}

The scalability issue in process mining is a big concern for applying the techniques on a large volume of data. 
Thus, different researchers investigated this problem through different techniques. 

Hern{\'a}ndez, S. et al. computed intermediate DFG and other matrixes through the MapReduce technique over a Hadoop cluster~\cite{HernHadoopPMIEEE2015}. 
The evaluation of their approach shows a similar trend for a performance like what we presented in \figurename{\ref{Fig:Experiment2}}. 
The performance cannot be compared precisely due to different setup and resources. 
This is the closest approach to ours. 
The advantage of our approach can be considered as the capability to preserve privacy while computing DFG. 

MapReduce has used by other researchers for the aim of process mining, e.g.,~\cite{reguieg2012using,evermann2014scalable}. 
As discussed by~\cite{HernHadoopPMIEEE2015}, MapReduce has used to support only event correlation discovery in~\cite{reguieg2012using}, and it is used to discover process models using Alpha Miner and the Flexible Heuristics Miner in ~\cite{evermann2014scalable}.

\subsection{Preserving privacy}

This paper does not directly contribute to preserving privacy issues in process mining; however, it enables new sort of support by enabling the use of fine-grain access controls, which are supported by Graph Databases, in particular Neo4j.

Preserving privacy in process mining is an important issue that recently got a lot of attention due to its importance. 
Mannhardt F. et al. describes guidelines for privacy in the process mining area~\cite{mannhardt2018privacy}. 
As technological privacy challenges, authors mentioned the \textit{aggregation challenge} that deals with enabling an analysis on aggregated event information without exposing the information at the individual level. 
The approach in this paper can well address this issue due to access control capabilities that exist in Neo4j. 
There are potentials to address other challenges using our approach, but the discussion is out of the scope of this paper.

Pika A. et al., proposed a framework for privacy-preserving when applying process mining in healthcare domain~\cite{pika2019towards}. 
They extend the work in ~\cite{pika2020privacy}, where they discussed the result of applying some techniques in other domains. 
As mentioned by authors, there might be needs for techniques like access control even after the data anonymization for output results. 
Our approach enables access control in one step before, i.e., DFG computation, and it does not address other issues in other steps.

There is also recent work to support the privacy issue in process mining like~\cite{bauer2019elpaas,michael2019user,fahrenkrog2019pretsa,mannhardt2019privacy,fahrenkrog2019providing}, but as the best of our knowledge, no approach applied access control and security when computing DFG.

\subsection{Graph database}

There are different attempts to use graph databases with process mining. Still, as our best of knowledge, none of the approaches supports shifting the calculation of DFG to a graph database. Thus, none of them can perform better than a traditional approach like pm4py.

Esser S. and Fahland D. used the graph database to query multi-dimensional aspects from event logs. 
This is one important use case that has been introduced by a graph database, i.e., adding more features to the data~\cite{esser2019storing}. 
They have used Neo4j as the graph database and used Cypher to query the logs. 
The approach uses a graph database as a log repository to sore data without any predefined structure, which is quite different from the topic of this paper. 
In this regard, the approach is similar to~\cite{de2016connecting}, where a relational database is used to store the data. 
The main difference is that ~\cite{esser2019storing} demonstrates that the graph database has more capability to add more features to data, which is a very important topic in any machine learning related approach in general.

Joishi J. and Sureka A. also used a graph database for storing non-structured event logs~\cite{joishi2016graph,joishi2015vishleshan}. 
They also demonstrated that Actor-activity matrix could be calculated using Cypher. 
However, the approach is context-dependent since the logs are not standardized like our approach. 
Also, the approach cannot be used with other process discovery algorithm since it does not shift and separate the computation of DFG to a graph database.

\section{Conclusion}\label{Sec:Conclusion}

This paper introduced and formalized a new approach to support process mining using graph databases. 
The approach defines how log files shall be stored in a graph database, and it also defined how Directly Follows Graph (DFG) can be calculated in the graph database. 
The approach is evaluated in comparison with pm4py by applying on a real log file.
The evaluation result shows that the approach supports mining processes when the event log is bigger than computational memory.
It also shows that it is scalable, and the performance is better when dicing the event log in a small chunk. 
The paper also discussed how this approach supports preserving privacy while computing DFG, which is an important topic in applying process mining in practice.

As future work, we intend to optimize our approach in terms of performance. 
Also, we intend to do some case studies to show how this approach supports privacy Preserving in practice. 
We also intend to develop a new library to support the use of a graph database for process mining for practitioners and researchers.

\bibliographystyle{plain}
\bibliography{References}

\begin{thebibliography}{10}

\bibitem{AalstBPMSurvey2013}
{W.M.P. van der} Aalst.
\newblock Business process management: a comprehensive survey.
\newblock {\em ISRN Software Engineering}, 2013, 2013.

\bibitem{AalstPMBook2016}
{W.M.P. van der} Aalst.
\newblock {\em Process Mining: Data Science in Action}.
\newblock Springer-Verlag Berlin Heidelberg, 2016.

\bibitem{AnglesGraphSurvey2008}
R.~Angles and C.~Gutierrez.
\newblock Survey of graph database models.
\newblock {\em ACM Computing Surveys (CSUR)}, 40(1):1--39, 2008.

\bibitem{bauer2019elpaas}
M.~Bauer, S.A. Fahrenkrog-Petersen, A.~Koschmider, F.~Mannhardt, H.~van~der Aa,
  and M.~Weidlich.
\newblock Elpaas: Event log privacy as a service.
\newblock {\em Proceedings of the Dissertation Award, Doctoral Consortium, and
  Demonstration Track at BPM}, pages 1--6, 2019.

\bibitem{berti2019process}
A.~Berti, S.J. van Zelst, and W.M.P. {van der}~Aalst.
\newblock {Process Mining for Python (PM4Py): Bridging the Gap Between
  Process-and Data Science}.
\newblock page 13–16, 2019.

\bibitem{WilCubeCase2015}
A.~Bolt, M.~De~Leoni, W.M.P. {van der}~Aalst, and P.~Gorissen.
\newblock Exploiting process cubes, analytic workflows and process mining for
  business process reporting: A case study in education.
\newblock In {\em SIMPDA}, pages 33--47, 2015.

\bibitem{de2016connecting}
E.G.L. De~Murillas, H.A. Reijers, and W.M.P. {van der}~Aalst.
\newblock Connecting databases with process mining: a meta model and toolset.
\newblock In {\em Enterprise, Business-Process and Information Systems
  Modeling}, pages 231--249. Springer, 2016.

\bibitem{dees2016bpi}
M.~Dees and B.F. van Dongen.
\newblock Bpi challenge 2016: Clicks not logged in.
\newblock 2016.

\bibitem{esser2019storing}
S.~Esser and D.~Fahland.
\newblock Storing and querying multi-dimensional process event logs using graph
  databases.
\newblock In {\em International Conference on Business Process Management},
  pages 632--644. Springer, 2019.

\bibitem{evermann2014scalable}
J.~Evermann.
\newblock Scalable process discovery using map-reduce.
\newblock {\em IEEE Transactions on Services Computing}, 9(3):469--481, 2014.

\bibitem{fahrenkrog2019providing}
Stephan~A Fahrenkrog-Petersen.
\newblock Providing privacy guarantees in process mining.
\newblock {\em Proceedings of the CAiSE Doctoral Consortium}, pages 23--30,
  2019.

\bibitem{fahrenkrog2019pretsa}
Stephan~A. Fahrenkrog-Petersen, H.~van~der Aa, and M.~Weidlich.
\newblock Pretsa: Event log sanitization for privacy-aware process discovery.
\newblock In {\em International Conference on Process Mining (ICPM)}, pages
  1--8. IEEE, 2019.

\bibitem{Cypher2018}
N.~Francis, A.~Green, P.~Guagliardo, L.~Libkin, T.~Lindaaker, V.~Marsault,
  S.~Plantikow, M.~Rydberg, P.~Selmer, and booktitle={International Conference
  on Management of Data} pages={1433--1445}~year={2018} Taylor, A.
\newblock Cypher: An evolving query language for property graphs.

\bibitem{HernHadoopPMIEEE2015}
S.~Hern{\'a}ndez, J.~Ezpeleta, S.J. {van}~Zelst, and W.M.P. {van der}~Aalst.
\newblock Assessing process discovery scalability in data intensive
  environments.
\newblock In {\em ACM 2nd International Symposium on Big Data Computing (BDC)},
  pages 99--104. IEEE, 2015.

\bibitem{JalaliBPIC2016}
A.~Jalali.
\newblock Exploring different aspects of users behaviours in the dutch
  autonomous administrative authority through process cubes.
\newblock {\em Business Process Intelligence (BPI) Challenge}, 2016.

\bibitem{joishi2015vishleshan}
J.~Joishi and A.~Sureka.
\newblock Vishleshan: performance comparison and programming process mining
  algorithms in graph-oriented and relational database query languages.
\newblock In {\em International Database Engineering \& Applications
  Symposium}, pages 192--197, 2015.

\bibitem{joishi2016graph}
J.~Joishi and A.~Sureka.
\newblock Graph or relational databases: A speed comparison for process mining
  algorithm.
\newblock {\em arXiv preprint arXiv:1701.00072}, 2016.

\bibitem{kimball2011data}
R.~Kimball and M.~Ross.
\newblock {\em The data warehouse toolkit: the complete guide to dimensional
  modeling}.
\newblock John Wiley \& Sons, 2011.

\bibitem{lenharthParallelGraph2016}
A.~Lenharth, D.~Nguyen, and K.~Pingali.
\newblock Parallel graph analytics.
\newblock {\em Communications of the ACM}, 59(5):78--87, 2016.

\bibitem{mannhardt2019privacy}
F.~Mannhardt, A.~Koschmider, N.~Baracaldo, M.~Weidlich, and J.~Michael.
\newblock Privacy-preserving process mining: Differential.
\newblock {\em Informatik Spektrum}, 42(5):349--351, 2019.

\bibitem{mannhardt2018privacy}
F.~Mannhardt, S.A. Petersen, and M.F. Oliveira.
\newblock Privacy challenges for process mining in human-centered industrial
  environments.
\newblock In {\em International Conference on Intelligent Environments (IE)},
  pages 64--71. IEEE, 2018.

\bibitem{michael2019user}
J.~Michael, A.~Koschmider, F.~Mannhardt, N.~Baracaldo, and B.~Rumpe.
\newblock User-centered and privacy-driven process mining system design.
\newblock In {\em CAiSE Forum}, volume 246, 2019.

\bibitem{pika2019towards}
A.~Pika, M.T. Wynn, S.~Budiono, A.H.M. ter Hofstede, W.M.P. {van der}~Aalst,
  and H.A. Reijers.
\newblock Towards privacy-preserving process mining in healthcare.
\newblock In {\em International Conference on Business Process Management},
  pages 483--495. Springer, 2019.

\bibitem{pika2020privacy}
A.~Pika, M.T. Wynn, S.~Budiono, A.H.M. ter Hofstede, W.M.P. {van der}~Aalst,
  and H.A. Reijers.
\newblock Privacy-preserving process mining in healthcare.
\newblock {\em International journal of environmental research and public
  health}, 17(5):1612, 2020.

\bibitem{reguieg2012using}
H.~Reguieg, F.~Toumani, H.R. Motahari-Nezhad, and B.~Benatallah.
\newblock Using mapreduce to scale events correlation discovery for business
  processes mining.
\newblock In {\em International Conference on Business Process Management},
  pages 279--284. Springer, 2012.

\bibitem{AalstReview2020}
W.M.P. {van der}~Aalst.
\newblock Academic view: Development of the process mining discipline.
\newblock In {\em Process Mining in Action: Principles, Use Cases and Outlook},
  pages 181--196. Springer, 2020.

\bibitem{WeskeBPM2019}
M.~Weske.
\newblock {\em Business process management: concepts, languages,
  architectures}.
\newblock Springer-Verlag Berlin Heidelberg, 2019.

\end{thebibliography}

\end{document}